\documentstyle[aps,12pt,pre,epsfig]{revtex}
\begin{document}

\draft
\tighten

\title{Energy flows in vibrated granular media}

\author{Sean McNamara$^{(1,2)}$ and Stefan Luding$^{(1)}$}
\address{(1) Institute for Computer Applications 1, Pfaffenwaldring 27,
70569 Stuttgart, GERMANY\\
(2) Benjamin Levich Institute and Department of Physics,
The City College of the City University of New York
New York, NY 10031, USA}

\date{\today}

\maketitle

\begin{abstract}

We study vibrated granular media, investigating each of the three
components of the energy flow: particle-particle dissipation,
energy input at the vibrating wall, and particle-wall dissipation.
Energy dissipated by interparticle collisions
is well estimated by existing theories
when the granular material is dilute, and these theories are
extended to include rotational kinetic energy.  
When the granular material is dense, the observed particle-particle 
dissipation rate decreases to as little as $2/5$ of
the theoretical prediction.
We observe that the rate of energy input is the weight of the granular
material times an average vibration velocity times a function of
the ratio of particle to vibration velocity.
`Particle-wall' dissipation has been neglected in all theories up to
now, but can play an important role when the granular material is
dilute.  
The ratio between gravitational potential energy and
kinetic energy can vary by as much as a factor of 3.  
Previous simulations and experiments have shown that $E\propto V^\delta$,
with $\delta=2$ for dilute granular material, and $\delta\approx1.5$
for dense granular material.
We relate this change in exponent to the departure of particle-particle
dissipation from its theoretical value.

\end{abstract}


\def\Et{\bar E}
\def\Er{\overcirc E}
\def\Eg{E_g}
\def\rp{r_p}
\def\rw{r_w}
\def\Bp{\beta_p}
\def\Bw{\beta_w}
\def\nt{\bar n}
\def\nr{\overcirc n}
\def\n{\bbox{\hat n}}
\def\bv{\bbox{v}}

\newpage

\section{Introduction}

\begin{center}
\begin{figure}[ht]
 \epsfig{file=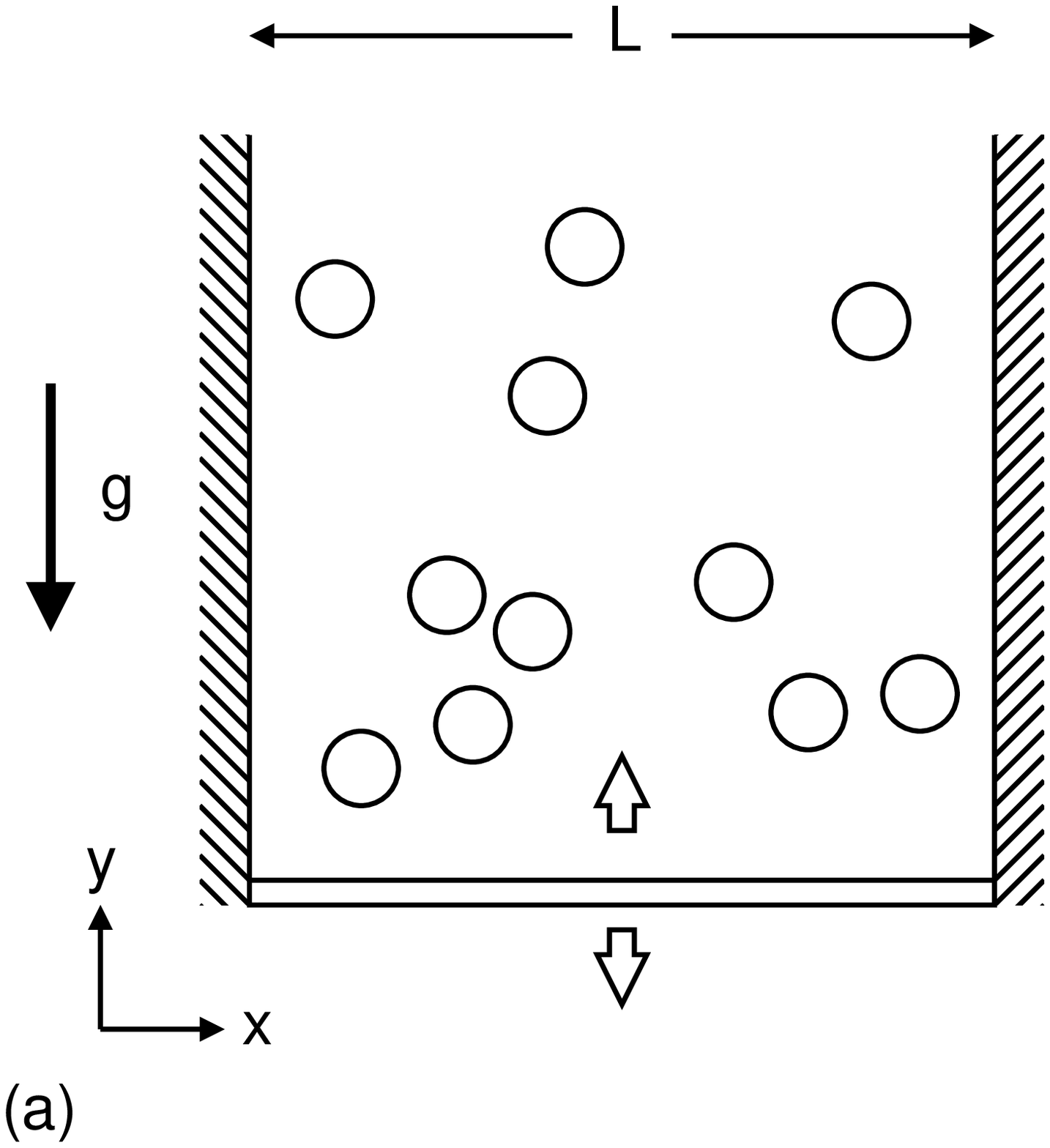,width=6.5cm,clip=,angle=0}
 \epsfig{file=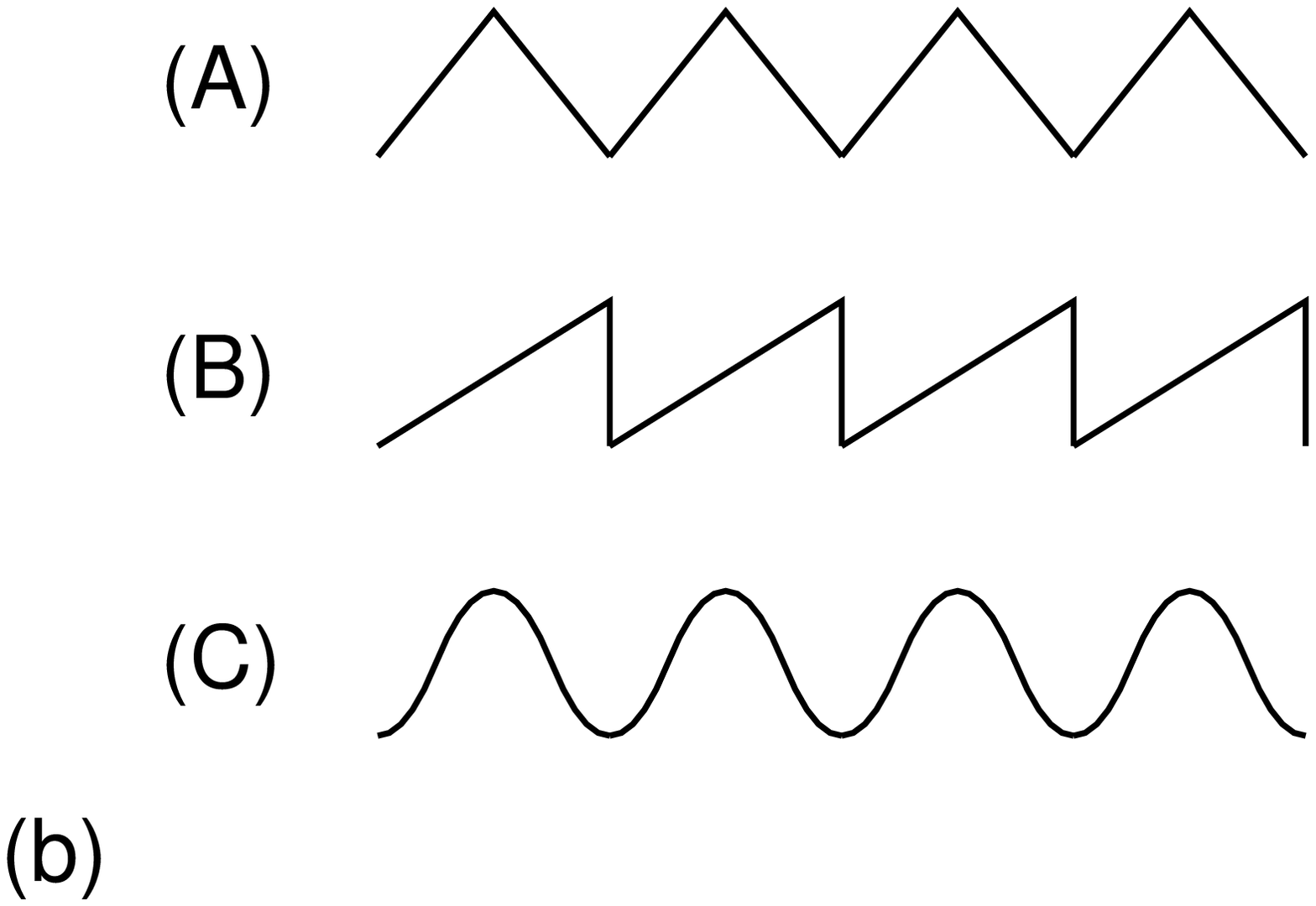,width=6.5cm,clip=,angle=0}
\caption{(a) A sketch of the studied system.  (b) The wave forms used
to drive the vibrating plate.}
\label{fig:system}
\end{figure}
\end{center}

In this paper, we study granular materials where  
energy is provided by a vibrating
plate  [see Fig.~\ref{fig:system}(a)].
In this system,
two general classes of motion can be imagined,
which we will name ``coherent'' and ``incoherent''.
In the first case, the particles move together in a coherent layer, bouncing
at some frequency related to the vibration frequency.  This state
exhibits surface waves, and has been the subject of much recent work
\cite{melo95,clement96,luding96e}.
Another possibility is that the motion is
incoherent.   The particles remain suspended above
the vibrating plate, with a density profile that -- excluding fluctuations --
remains constant
throughout the vibration period.  This second state has also been the
subject of recent experiments, simulations and theories
\cite{rosato93,luding95b,lee95,warr95,goldshtein95b,kumaran,muller97}.

In this paper, we study this second type of motion.  We search for
a basic understanding of the physical processes which govern the input and
dissipation of energy.  This is a question of theoretical interest,
because it provides a good test of kinetic theories of granular
media.  These theories, modeled after the kinetic theory
of gases\cite{jenkins85,lun91,goldshtein95}
have been shown to have quantitative success only
for unforced granular media
in the absence of gravity \cite{lun91,goldhirsch93,mcnamara96,deltour97}.
However, incoherent vibrated granular
materials may be an experimentally accessible system well described 
by these theories.\cite{warr95,goldshtein95b,goldshtein95}

The paper is organized as follows.  In Sec.~\ref{sec:system}, we
describe the studied system in detail and set forth our notation.
In Sec.~\ref{sec:review}, we review the literature.  In
Sec.~\ref{sec:simulations}, we check the previous results against
our simulations, and construct some new theories.  In particular,
we examine five topics: 1) the rate of energy dissipation in
particle-particle collisions, 2) the effect of particle rotations,
3) the energy input by the vibrating floor, 4) the energy dissipated
by particle-wall collisions, and 5) the ratio of kinetic to
gravitational potential energy.  Finally, in Sec.~\ref{sec:grail},
we assemble the best theories and investigate the dependence of 
the energy on vibration velocity.

\section{Description of the system}
\label{sec:system}

\subsection{Definitions}

A sketch of the system is shown in Fig.~\ref{fig:system}(a).
A granular medium, modeled
by a gas of inelastic disks, is contained in a box of width $L$
and infinite height.  The gas is pushed against the bottom by a
gravitational field with acceleration $g$.  Energy is added to the system
by the bottom of the box, which vibrates with period $\tau$ and typical 
velocity $V$.  (In Sec.~\ref{sec:Pb}, we relate $V$ to the maximum 
velocity $V_{\text{max}}$.)
The granular medium consists of $N$ particles of radius $a$ and
mass $m$.  Convenient nondimensional
parameters describing the system are the length measured in particle radii,
the number of layers of particles at rest $H\equiv 2aN/L$,
and the number
of particle radii a particle, initially at rest, falls during one
cycle of a wall vibration, $g\tau^2/(2a)$.  The description of the
system is completed by specifying the particle-particle and the
particle-wall collisions, and the wave form of the vibrating wall.
We use the three wave forms shown in Fig.~\ref{fig:system}(b).
Wave forms $A$ and $B$ are convenient for theoretical analysis 
\cite{warr95,kumaran,mcnamara97}, and wave form $C$ 
is a computational convenient numerical approximation to a sine wave,
constructed by patching together parabolas.  In Sec.~\ref{sec:Pb}, we
will estimate the effect of this approximation on the energy input at
the vibrating wall.

We consider collisions with constant
normal and tangential restitution coefficients.  
This collision model has
been widely studied and is relatively easy to analyze theoretically.
The normal (tangential) restitution coefficient for collisions
between particles is denoted $\rp$ ($\Bp$), and for collisions
between the particles and the side walls $\rw$ ($\Bw$).  The
vibrating floor is elastic ($r=1$) and smooth ($\beta=-1$).  More
details on the collision rule are presented in
Appendix~\ref{sec:Collrule}. 

We describe the state of the system with three types of energy,
each defined per degree of
freedom, per particle: $\Et$, the
translational kinetic energy of the particles, $\Er$, their rotational
energy, and $\Eg$, their gravitational potential energy.  These energies
are calculated as follows:
\begin{equation}
\Et = {m\over2 N\nt} \sum_{i=1}^N v_i^2, \quad
\Er = {mqa^2\over2 N\nr} \sum_{i=1}^N \omega_i^2, \quad
\Eg = mg{1\over N} \sum_{i=1}^N (y_i-h_0).
\end{equation}
Here, $v_i$ is the translational velocity of the $i^{\text{th}}$
particle, $\omega_i$ is its angular velocity, and $y_i$ is
its height above the time averaged position of the vibrating floor.  
The center of mass of the particles at rest is $h_0$.  We calculate
$h_0$ from Eq.~(14) of Ref.~\cite{luding95b}.  The number
of translational (rotational) degrees of freedom per particle is $\nt$
($\nr$).  In this paper, we consider
exclusively two dimensional systems, where $\nt=2$ and $\nr=1$.
$\Et$ and $\Er$ are often called ``granular temperatures''. 
This terminology is not meant to imply that a thermal equilibrium
exists in granular flows, but simply to draw an analogy between these
quantities and the temperature of an ideal gas, which is proportional
to the average energy per degree of freedom.

\subsection{General Approach}

The ``holy grail'' of this paper are expressions for the energies
$\Et$, $\Er$, and $\Eg$ in terms of the parameters $L$, $g$, $V$,
$a$, $\tau$, $m$, $\rp$, $\rw$, $\Bp$ and $\Bw$,
together with a physical understanding of the system.
The starting point of the quest is the equation for the energy
flow at steady state:
\begin{equation}
P_b = D_{pp} + D_{pw},
\label{eq:EnergyBalance}
\end{equation}
where $P_b$ is the power input by the bottom, $D_{pp}$ is the power
dissipated by particle-particle collisions, and $D_{pw}$ is the power
dissipated by particle-wall collisions.  We will try to express
each of these quantities in terms of $\Et$ and the system
parameters.  Then Eq.~(\ref{eq:EnergyBalance})
will give an expression for $\Et$, and we will also understand
what processes determine $\Et$.

In simulations, it is possible to monitor each of the three quantities
in Eq.~(\ref{eq:EnergyBalance}) or set each of them either to zero or to
a known constant.  $D_{pp}$ can be set to $0$
by using elastic particles, just as using elastic walls permits one to
set $D_{pw}=0$.  $P_b$ can be set to a known value by driving the
bottom wall with waveform $B$\cite{mcnamara97}.  We will discuss each of
the three terms in Eq.~(\ref{eq:EnergyBalance}) and exploit
this separability to construct and verify our theories piece by piece.  

\subsection{Simulational Approach}

We use the standard event-driven method,
which has been previously used to study vibrated
granular materials \cite{luding96e,luding95b,luding94c,luding97c}.  
Its main approximation
is to consider that collisions are instantaneous, i.e. the time of
contact during a collision is $0$.  The main disadvantage of
event-driven simulations for granular materials is the occurrence of
inelastic collapse -- an infinite number of collisions occurring in
a finite time \cite{bernu90,mcnamara92}.  To circumvent this problem,
dissipation is turned off when the inelastic collapse singularity is
approached.  A similar method has been used in other recent work
\cite{luding96e,luding97c}.  The fraction of dissipationless collisions is
always less than $10^{-4}$ for the simulations shown here.  In addition, we
repeated the three simulations that had the most dissipationless collisions,
approaching more closely the inelastic collapse singularity, and found no
measurable change

In the simulations presented here, we use the particle radius to define
the unit of distance, and the particle mass to define the unit of mass.
The unit of time is arbitrary.  To apply our results to experimental
systems, it is necessary to introduce conversion factors based on
the particle mass and the particle radius, and to choose $\tau$ and $g$ so
that $g\tau^2/a$ has the same value as the experimental system.

In order to vary the system parameters in a consistent and organized
way, the following procedure was used.  First, `central' values
were selected for all the parameters except $V$.  Then $V$ is swept
from small values to very large values, care being taken that all
simulations fall within the ``incoherent'' category discussed above.
The central values used in this paper are $N=160$, $L=50$, $\rp=0.95$
$\rw=1$, $g=1.0$ and $\tau=1.0$.  Then
for each parameter, two series of simulations are run, one where
the parameter is increased by a factor of $5$, and another where it
is decreased by a factor of $5$.  There are two exceptions: $1-\rp$
(not $\rp$) is increased or decreased by a factor of 5, and
$g$ is increased or decreased by a factor of 25.

Most of the quantities measured during the simulations must be averaged
over long periods of time to obtain stable results.  To obtain a suitable
averaging time, an ``energy turn--over time'' $\tau_E = \bar E/P_b$ was
estimated.  The averaging time was taken to be $20\tau_E$.  It was verified
that the actual $\tau_E$ was always close to the estimated one.  Initial
transients by running each simulation over several averaging periods,
and verifying that averages were not changing significantly with time.

\section{Review of Previous Work}
\label{sec:review}

\subsection{Experiments and theory of Warr, et al.}
\label{sec:Warr}
Warr, Huntley and Jacques\cite{warr95} modeled vibrated granular media
as an isothermal ``atmosphere''.  By integrating over the ``atmosphere'',
one obtains
\begin{equation}
D_{pp} = 2(1-\rp^2) NmgH (\Et/m)^{1/2}.
\label{eq:WarrDpp}
\end{equation}
Next, one can estimate the energy added by considering the density
at the bottom of the ``atmosphere''.  The result for wave form $A$ is
\begin{equation}
P_b = (1/2)NmgV^2 (m/\Et)^{1/2},
\label{eq:WarrPw}
\end{equation}
Setting $P_b=D_{pp}$ yields 
\begin{equation}
\Et = {mV^2 \over 4H(1-\rp^2)}.
\label{eq:Warrtheory}
\end{equation}
This result was calculated for smooth particles ($\Bp=-1$), with
the system being driven by wave form $A$. 
The authors also present experimental results suggesting
\begin{equation}
\Et \sim H^{-\bar\nu} V^{\bar\delta} \quad
\Eg \sim mg H^{-\nu_g} V^{\delta_g}.
\label{eq:Warrexp}
\end{equation}
with $\bar\nu = 0.6\pm0.03$, $\bar\delta = 1.41\pm0.03$, $\nu_g = 0.27\pm0.11$
and $\delta_g=1.3\pm0.04$.  In contrast, the theory assumes that the
gravitational potential energy is always proportional to the kinetic
energy with $\bar\nu=\nu_g=1$ and $\bar\delta=\delta_g=2$.
Explaining the discrepancy between Eq.~(\ref{eq:Warrtheory})
and Eq.~(\ref{eq:Warrexp}) was one of the major motivations for our study.

\subsection{Simulations of Luding}
\label{sec:Luding2D}
Luding and co-workers have studied this problem with numerical
simulations.  Simulations without rotation \cite{luding94c}
give $\nu_g\approx1$, $\delta_g\approx1.5$.  Simulations
including rotation, carefully planned to duplicate the experiments
\cite{luding95b} give $\nu_g=0.76\pm0.11$ and $\delta_g=1.60\pm 0.10$.  
When $V$ is made very large and dissipation at the
walls is suppressed \cite{luding95b,muller97}, $\delta\to2$.  
Lee \cite{lee95} obtained results that are qualitatively similar.
Luding \cite{luding94c} used a slightly different collision model which
accounts for Coulomb friction.  In the limit of infinitely strong
friction, it reduces to the model used in this paper.

\subsection{Theory of Kumaran}
\label{sec:Kumaran}
Kumaran\cite{kumaran} followed the same approach as Warr,
Huntley and Jacques\cite{warr95}, except that he assumes that the
particle velocities are distributed according to a Maxwellian velocity
distribution.  On the other hand, Warr, et.~al assumed that all particles
possess the mean velocity.  Thus, Kumaran obtains the same results
as Warr, et.~al, except for the constant prefactor.  Kumaran also
investigates both wave forms $A$ and $B$.  He obtains
\begin{mathletters}
\label{eq:kumaran}
\begin{equation}
D_{pp} = \sqrt{\pi/2}(1-\rp^2)NmgH (\Et/m)^{1/2},\quad\text{and}
\label{eq:KumaranDpp}
\end{equation}
\begin{equation}
P_b = Nmg \left( \langle V \rangle + 
   2\sqrt{m\over\pi \Et} \langle V^2 \rangle + O(mV^3) \right),
\label{eq:KumaranPw}
\end{equation}
\end{mathletters}
where the angle brackets indicate an average taken over one period.
If a wave form is symmetric, like wave forms $A$ and $C$, then
$\langle V \rangle = 0$, and $P_b \sim N mgV^2(m/\Et)^{1/2}$, 
as in Eq.~(\ref{eq:WarrPw}), but
when the wave form is asymmetric (wave form $B$), then the 
first term of the series is nonzero and dominant, so that $P_b \sim NmgV$.

\subsection{Theory and Simulations of McNamara and Barrat}
\label{sec:JeanLouis}
McNamara and Barrat\cite{mcnamara97}
studied the input of energy at a vibrating wall in
the absence of gravity.  They also found a difference between wave form
$B$ and wave forms $A$ and $C$.  For wave form $B$, the power
input by the vibration wall is $P_b=pVL$, where
$p$ is the average pressure on the vibrating wall.  This equation can
be derived by considering the encounter of a single particle with
a moving wall at time $t_*$.  The particle's change in energy $\Delta E$
is related to its change in momentum $\Delta p$ by 
$\Delta E = V(t_*) \Delta p$, where $V(t_*)$ is the wall velocity at
time $t_*$.
For wave form $B$, $V(t_*)=V$ always,
so it is easy to average over time and find  that the energy input is $pVL$.  
For wave forms $A$ and $C$, the velocity of the plate is not always
the same when the particles hit the plate.  However, the probability
of a particle hitting the plate at a given phase is governed only
by the quantity $U/V$, where $U$ is the velocity of the incoming particle.
Therefore, for wave forms $A$ and $C$,
$P_b=pVL f(U/V)$, where the function $f$ is unknown.  
To adopt these results to a system under
gravity, one sets $pL=Nmg$, the weight of the granular material.
The result for wave form $B$,
\begin{equation}
P_b = NmgV,
\label{eq:NmgV}
\end{equation}
agrees with Eq.~(\ref{eq:KumaranPw}) for asymmetric wave forms.  For
wave forms $A$ and $C$, $P_b=NmgV f(U/V)$, which coincides with
Eqs.~(\ref{eq:WarrPw}) and~(\ref{eq:KumaranPw}) if $f(U/V)=V/U$.

\section{Results}
\label{sec:simulations}
\subsection{Particle-Particle dissipation of smooth particles}
\label{sec:Dpp}
\subsubsection{Simulations}

\begin{center}
\begin{figure}[ht]
 \epsfig{file=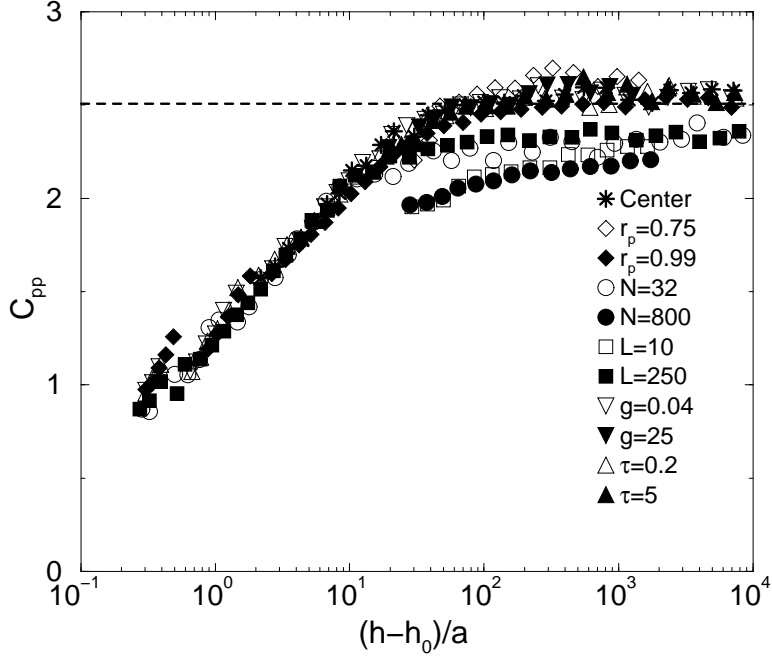,width=12cm,clip=,angle=0}
\caption{A test of the scalings in Eqs.~(\ref{eq:WarrDpp}) 
and~(\ref{eq:KumaranDpp}).  
Here, $C_{pp}$ is plotted against
the height of the center of mass $h\equiv N^{-1} \sum y_i$ above
its position at zero energy, $h_0$.
Both equations predict that $C_{pp}$
should be a constant, which is indeed true for large heights (dilute
granular media).  The central simulation has $N=160$, $\rp=0.95$,
$L=50$, $g=1$, and $\tau=1$; $V$ is varied over several orders of magnitude.
The other series of simulations, each indicated by a different symbol,
have the same parameters as the central
one, except for the parameter shown in the legend.  The bottom is
driven with wave form $B$, and the side walls are replaced by
periodic boundary conditions.  The dashed line
shows the constant predicted by Eq.~(\ref{eq:KumaranDpp}).}
\label{fig:Dpp}
\end{figure}
\end{center}

The two expressions for $D_{pp}$ in Eq.~(\ref{eq:WarrDpp}) and 
Eq.~(\ref{eq:KumaranDpp}) differ only by a constant.
We measure this constant by calculating 
\begin{equation}
C_{pp} \equiv {D_{pp} \over (1-\rp) NmgH (\Et/m)^{1/2}}.
\label{eq:defCpp}
\end{equation}
Eq.~(\ref{eq:WarrDpp}) predicts $C_{pp} = 4$
and Eq.~(\ref{eq:KumaranDpp}) predicts $C_{pp} = \sqrt{2\pi}
\approx 2.5$.  [The factor of $2$ difference between the predicted
values $C_{pp}$ and Eqs.~(\ref{eq:WarrDpp}) and~(\ref{eq:KumaranDpp})
arises from replacing $1-\rp^2$ with $2(1-\rp)$.
For $\rp\approx1$, $1-\rp^2 \approx 2(1-\rp)$,
so one might think that these two quantities would be equivalent.
However, the difference at $\rp=0.75$ is sufficiently
great to show that $1-\rp$ collapses the data better than $1-\rp^2$.]
Fig.~\ref{fig:Dpp} shows that the scaling successfully collapses the
simulation data onto a single curve.
The figure can be divided into two regions: the dilute region, $h-h_0>50a$,
and the dense region, $h-h_0<50a$.  In the dilute region, $C_{pp}$
is a constant, and quite close to the value predicted by
Eq.~(\ref{eq:KumaranDpp}).   The points collapse into families determined
by the particular value of $H=2aN/L$ of each simulation.  However,
the scaling captures the dependence on $V$, $\rp$, $g$ and $\tau$ to
within the noise in the simulations.  In the dense region, $h-h_0<50a$,
{\sl all} simulations
collapse tightly onto a curve $C_{pp} \sim \log[(h-h_0)/a]$.  
The reason for this dependence of $C_{pp}$ on $(h-h_0)/a$ is unknown.
It is not due to the increasing density, because the low $H$
($N=32$ and $L=250$ curves in Fig.~\ref{fig:Dpp}), as well as
additional simulations (not shown) also collapse
onto the same curve.  Nor is the deviation due to waves propagating
upwards from the vibrating plate; the agreemt of the $\tau=0.2$ curve
with the rest of the simulations excludes this.  We also tried an
alternate way of adding energy: the bottom plate is held fixed, and
when a particle hits the bottom, it is given a velocity drawn from
a Maxwellian distribution.  But these simulations also reproduce the
curve shown in Fig.~\ref{fig:Dpp}.  Replacing periodic boundaries
with elastic walls modifies only slightly Fig.~\ref{fig:Dpp}.  

Fitting a straight line to the points with $a<h-h_0<50a$ (and
excluding the $N=800$ and $L=10$ points) gives 
$C_{pp} \approx 0.30 \log[(h-h_0)/a] + 1.35$.  
This relation is
valid over only about one and half orders of magnitude, so it is 
difficult to say whether it is really logarithmic, or whether the
logarithm is only a convenient approximation. 

\subsubsection{A physical argument}
\label{sec:Hafflike}
We now derive the scaling relation from physical arguments.  This
permits us not only to understand why $1-\rp$ is better than $1-\rp^2$,
but also enables us to extend the theories to account for particle
rotations and dissipation at the side walls.

From general kinetic theory arguments in the style
of Haff \cite{haff83}, we argue that the dissipation due to
collisions between particles will be
\begin{equation}
D_{pp} \sim N \langle\Delta E\rangle \left({U\over s}\right),
\label{eq:Dppscaling}
\end{equation}
where $\langle\Delta E\rangle$ is the average energy lost per collision,
$U$ is a typical particle velocity, and $s$ is a typical
particle separation. 
Multiplying $\langle\Delta E\rangle$ by the collision frequency
$U/s$ gives the energy dissipation per particle per unit time.
Then multiplying
by the number of particles gives the total energy dissipation in the
system.
 
Next, we must relate $\langle\Delta E\rangle$, $U$, and $s$
to $\Et$ and the independent parameters.
First of all, Eq.~(\ref{eq:DeltaEappendix}) in the appendix shows that
$\langle\Delta E\rangle\sim(1-\rp^2)\Et$ for smooth ($\Bp=-1$)
particles.  Next,
since $U$ is a typical velocity, we make the identification
$U \sim (\Et/m)^{1/2}$.  Finally, to estimate the particle separation $s$,
we turn to a one dimensional model.  We imagine a column of $n$
particles, suspended in a gravitational field by the vibrating plate at
the bottom of the column.  The column is in a steady state, so the
net force due to collisions on each particle, averaged over time, must 
balance the gravitational force $mg$.  Consider
a particle somewhere in the column, with $M$ particles above it
($1<M<n$).  In a steady state, this particle must receive a momentum
flux $(M+1)mg$ from the particle below it and transfer $Mmg$ to the particle
above it.  The momentum transferred per
collision is $(1+\rp)mU$, and the collision
frequency is $U/s$, hence the momentum flux will be proportional to
$m(1+\rp)U^2/s$.  Equating the two expressions for the momentum flux
\begin{equation}
Mmg \sim (1+\rp)mU^2/s,
\end{equation}
and solving for $s$ gives $s \sim (1+\rp)U^2/(Mg)$.
$M$ will scale as $n$, and in two dimensions $n$ can be 
approximated by the number of layers of particles $H=2Na/L$.  Again using
$U^2\sim \Et/m$, we have $s \sim (1+\rp)\Et /(mgH)$.
 
Inserting our expressions for $s$ and $U$ into Eq.~(\ref{eq:Dppscaling}),
we have
\begin{equation}
D_{pp} = C_{pp} (1-\rp) NmgH (\Et/m)^{1/2}.
\label{eq:Dpp}
\end{equation}
This agrees in order of magnitude with the scaling tested in
Fig.~\ref{fig:Dpp}, and, in addition, explains why the $1-\rp$ of
Eq.~(\ref{eq:defCpp}) collapses the data better than the $1-\rp^2$
of Eqs.~(\ref{eq:WarrDpp}) and~(\ref{eq:KumaranDpp}).

\subsection{Particle-particle dissipation for rough particles}
\label{sec:rotate}
 
All the simulations and theories presented so far in this paper consider
``smooth'' particles ($\Bp=-1$), thus ignoring rotation.  But
particles in granular flows rotate, and our theory would not be
complete without considering rotation.  

\begin{center}
\begin{figure}[ht]
 \epsfig{file=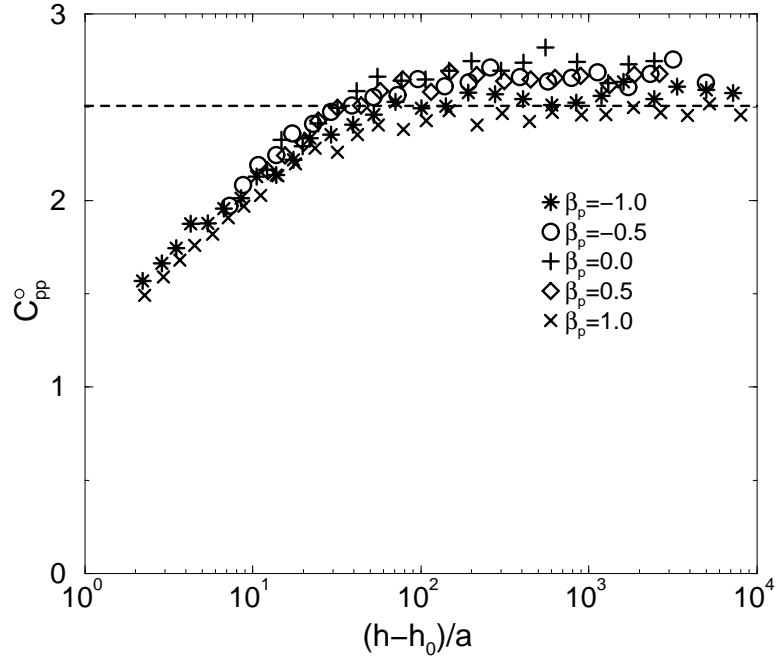,width=12cm,clip=,angle=0}
\caption{A test of the scaling in Eq.~(\ref{eq:Dpprotate}), which
defines $C^{\circ}_{pp}$.  The parameters for the simulations are
$N=160$, $L=50$, $g=1$, $\tau=1$, and
$\rp=0.95$, with $\Bp$ given in the figure and $V$
varied between $100$ and $0.35$.  The dotted line and the $\Bp=-1$
simulation also appear in Fig.~\ref{fig:Dpp}.  Wave form $B$ is used,
and the side walls are replaced by periodic boundaries.}
\label{fig:Dppbeta}
\end{figure}
\end{center}

We now revise our argument for the scaling of $D_{pp}$ in
Sec.~\ref{sec:Dpp}, to take into account rotation.  The only place
the smoothness of the particles entered was in estimating 
the change in energy per collision $\langle\Delta E\rangle$
in Eq.~(\ref{eq:Dppscaling}), so we use the more complicated
expression given in the appendix, Eq.~(\ref{eq:DeltaEappendix}):
\begin{equation}
\langle\Delta E\rangle = - 2 (1-\rp^2) \Et
		- {q(1-\Bp^2)\over 1+q}\left(\Et+\Er/q\right).
\label{eq:DeltaE}
\end{equation}
This involves $\Er$ as well as $\Et$, so we draw from a recent
paper\cite{mcnamara97b} on the ratio of translational
to rotational kinetic energy in granular flows.  One of the principle
results of this paper is that
\begin{equation}
K \equiv {\Et\over\Er} = {1+2q-\Bp \over q(1+\Bp)},
\end{equation}
in vibrated granular flows.  This convenient fact enables us to
replace $\Er$ with $\Et/K$.  The result is
\begin{equation}
\langle\Delta E\rangle = - 2 \left[(1-\rp^2)
		+ {q(1-\Bp^2) \over 1+2q-\Bp} \right] \Et
	= - 2 [1-{\rp^*}^2] \Et.
\end{equation}
with the ``effective restitution coefficient'', governing
the loss of energy,
\begin{equation}
{\rp^*}^2 = \rp^2 - {q(1-\Bp^2) \over 1+2q-\Bp}.
\label{eq:reff}
\end{equation}
Inserting $\langle\Delta E\rangle \sim (1-{\rp^*}^2) \Et$ into
Eq.~(\ref{eq:Dppscaling}) gives
\begin{equation}
D_{pp} = C^{\circ}_{pp} \left( {1-{\rp^*}^2 \over 1+\rp} \right)
		 NmgH (\Et/m)^{1/2}.
\label{eq:Dpprotate}
\end{equation}
as the analogy of Eq.~(\ref{eq:Dpp}).  This equation also defines a
constant $C^{\circ}_{pp}$, in analogy with $C_{pp}$.  We plot
$C^{\circ}_{pp}$ in Fig.~\ref{fig:Dppbeta}.
There remains a weak dependence on $\Bp$, but $C_{pp} = C^{\circ}_{pp}$
within the accuracy of the simulations.

\subsection{Energy input}
\label{sec:Pb}

We now turn our attention to the energy added by the vibrating floor.
We test first the simplest result: Eq.~(\ref{eq:NmgV}), which applies only
to wave form $B$.  Reviewing all simulations shown in Fig.~\ref{fig:Dpp},
we find the largest deviation from Eq.~(\ref{eq:NmgV}) is $1.6$\%, with
most others much less ($90$\% of the particles deviate by $0.5$\% or less). 
This equation is accurate because it is independent of the velocity
distribution of the particles, and can be derived from the conservation
of momentum\cite{mcnamara97}.  

\begin{center}
\begin{figure}[ht]
 \epsfig{file=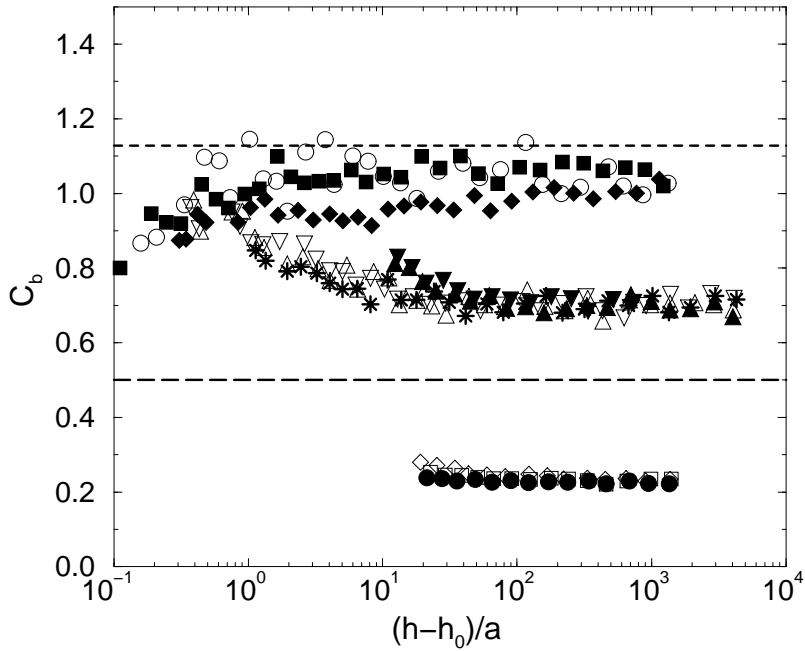,width=12cm,clip=,angle=0}
\caption{A test of Eqs.~(\ref{eq:WarrPw}) and (\ref{eq:KumaranDpp}):
we plot $C_b \equiv P_b / [NmgV^2(m/\Et)^{1/2}]$
for the symmetric wave form $A$.
The symbols and parameters of the simulations are the same as in
Fig.~\ref{fig:Dpp}, except the bottom plate is driven by wave form $A$
instead of wave form $B$.  The dashed lines show the constants predicted
by Eqs.~(\ref{eq:WarrPw}) and~(\ref{eq:KumaranPw}).}
\label{fig:Pw2}
\end{figure}
\end{center}

For wave form $A$, Refs.~\cite{warr95,kumaran} predict
$P_b \sim NmgV^2(m/\Et)^{1/2}$.
We test this scaling against the simulations in Fig.~\ref{fig:Pw2}.
We see that the scaling is only
partially successful.  The rescaled power input varies by as much as a
factor of $5$, and there is a strong dependence on $H(1-\rp)$ not
captured by the scaling.  An analogous plot for wave form $C$ is similar.

To find a better way to calculate $P_b$ for wave forms $A$ and
$C$, we try the scaling suggested by
Ref.~\cite{mcnamara97}: $P_b = NmgV f(U/V)$ with
$U = (\Et/m)^{1/2}$.  Let us now consider the unknown function
$f(U/V)$ in the limits $U/V\to\infty$ and $U/V\to0$.  In the first limit,
the particles are moving infinitely more quickly than the wall, and
thus have an equal probability of hitting the wall when it is ascending
or descending.  In this case, the net energy input by the wall will be
$0$ because the energy gained by particles during the ascending phase
is lost during the descending phase.  As $U$ becomes smaller, the particles
have a higher probability to hit the wall during its ascending phase,
and hence $P_b$ becomes positive.  Finally, in the limit $U\to0$,
the particles can hit the wall only during the ascending phase, so
wave form $A$ becomes equivalent to wave form $B$, so
$f(U/V)\to1$.  This limiting value of $f$ is also attained for wave
form $C$ if $V=2V_{\text{max}}/3$.  In all cases, $V$ is equal
to a space average of the velocity of the plate during its ascending
phase, $V=A^{-1} \int_0^A V(y) dy$, where $y=A$ is the maximum height
attained by the plate.  When the plate sweeps through a motionless gas
of particles, $V$ is the average plate velocity seen by the particles
(assuming all particles collide only once with the plate).
For the sinusoidal wave form, a similar calculation gives
$V = \pi V_{\text{max}} /4$.

\begin{center}
\begin{figure}[ht]
 \epsfig{file=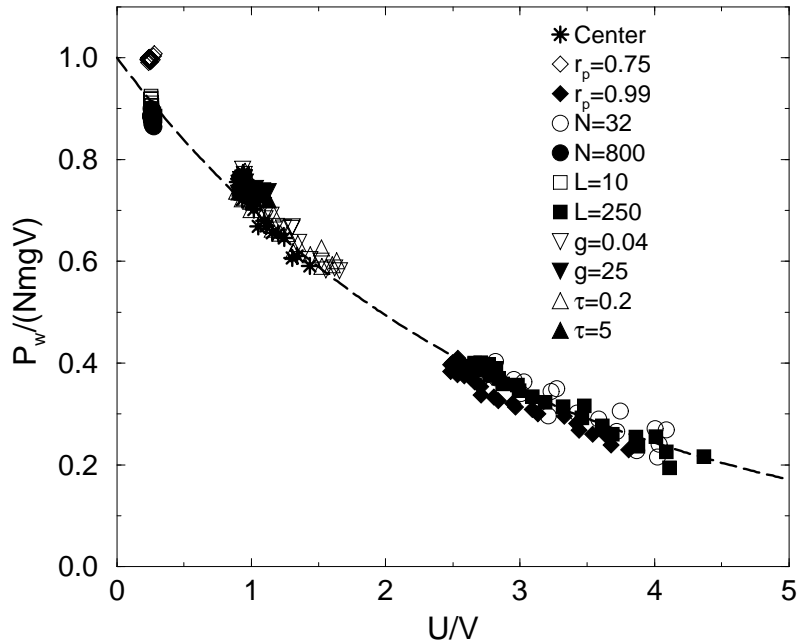,width=12cm,clip=,angle=0}
\caption{A graph of $f(U/V) = P_b/(NmgV)$, with
$U=(\Et/m)^{1/2}$.  The parameters and symbols of the simulations
are the same as in Fig.~\ref{fig:Dpp}, except that wave form
$A$ is used. }
\label{fig:eff}
\end{figure}
\end{center}

The unknown function $f$ is shown
in Fig.~\ref{fig:eff} for wave form $A$.  An analogous plot for
wave form $C$ is similar.
For wave forms $A$ and $C$, points fall on a single curve, 
but they are bunched together
depending on the value of $(1-\rp)H$.  The function $f$ is well
approximated by an exponential, so that
\begin{equation}
P_b = NmgV \exp [-\alpha U/V].
\label{eq:Pwall}
\end{equation}
Least squares fits give $\alpha_A=0.353\pm 0.002$, and
$\alpha_C=0.475\pm0.003$.  (Of course, $\alpha_B=0$.)
We note that $\alpha_A / \alpha_C \approx 3/4$.
Fig.~\ref{fig:eff} does not prove that $f$
is an exponential, because we have calculated $f$ over less than
one order of magnitude.  However, an exponential is a better
approximation for $f$ than a power law or a straight line.

Note that in all scalings, $P_b$ depends on the plate motion only
through $V$ and not through $\tau$.  This differs from previous
simulations, which suggest that $P_b$ drops dramatically when
$\tau$ is below a critical value \cite{rosato93}.  However, 
Ref.~\cite{rosato93} used a soft sphere simulation method, where particles
remain in contact during a finite time while colliding.  If
$\tau$ approaches this contact time, the efficiency of the forcing
will drop.   On the other hand, if $\tau$ is made very large, the
granular material dissipates most of its energy during one vibration
cycle, and the transition to the ``coherent'' state occurs.  In the
coherent state, the relevant parameter describing the forcing is no
longer $V$, but the acceleration $V/\tau$.

\subsection{Effect of Side Walls}
\label{sec:walls}

\begin{center}
\begin{figure}[ht]
 \epsfig{file=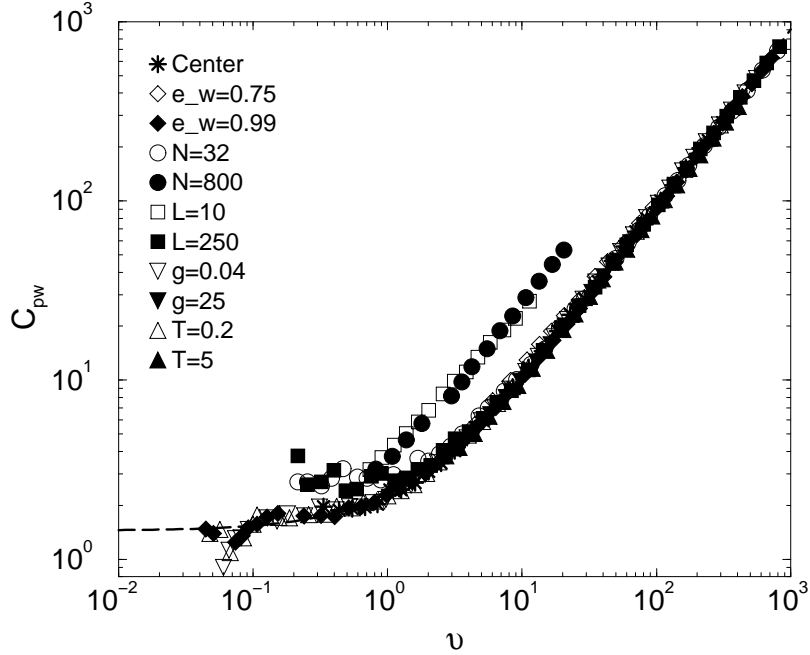,width=12cm,clip=,angle=0}
\caption{A test of the scaling in Eq.~(\ref{eq:Dpw}), where $C_{pw} \equiv
(D_{pw}/D_{pp})(L/a)(\bar E/\bar E_x)^{3/2}$ is linear in 
$\upsilon\equiv(h-h_o)/(aH)=L(h-h_0)/(2Na^2)$.
The dotted line is $C_{pw}=1.45 + 0.905\upsilon$.  These values were
obtained by fitting all the points on the graph except those of the
large $H$ simulations ($L=10$ and $N=800$).  These simulations have the
same parameter values as those in Figs.~\ref{fig:Dpp},~\ref{fig:Pw2}
and~\ref{fig:eff}, except that periodic boundaries have been replaced
by walls with $r_w=r_p$ and $\beta_w=-1$.}
\label{fig:Dpw}
\end{figure}
\end{center}

Finally, we consider dissipation of energy through collisions between
the particles and the side walls.
There is no theory for $D_{pw}$ in the literature.  Nevertheless, we find
that $D_{pw}$ can be estimated by
\begin{equation}
D_{pw} = C_{pw}(\upsilon) D_{pp} \left({a\over L}\right) 
	\left( {\bar E_x \over \bar E}\right)^{3/2},
\label{eq:Dpw}
\end{equation}
where $\upsilon\equiv(h-h_o)/(aH)=L(h-h_0)/(2Na^2)$ is proportional
to the specific volume or inverse density.
$C_{pw}(\upsilon)$ is a linear function, and $\bar E_x$
is the kinetic energy per degree of freedom
in the $x$ component of the particle velocities:
$\bar E_x = m(2N)^{-1} \sum v_{xi}^2$.
In a gas at thermal equilibrium, $\bar E_x/\bar E = 1$, but in these
simulations, $\bar E_x/\bar E$ can attain values as high as $0.2$.
The function $C_{pw}(\upsilon)$ is shown with the data from the
simulations in Fig.~\ref{fig:Dpw}.
The simulations agree well with Eq.~(\ref{eq:Dpw}), except for the large
$H$ ($L=10$ and $N=800$) simulations. 

The result Eq.~(\ref{eq:Dpw}) can be understood as a continuation between
two limits.  In the limit of $\upsilon\ll1$, Eq.~(\ref{eq:Dpw}) becomes
$D_{pw} \sim D_{pp} (a/L) (\bar E_x/\bar E) ^{3/2}$.  The factor of
$(\bar E_x/\bar E) ^{3/2}$ arise the $x$ velocities alone determine the
frequency of collisions with the wall, and the amount of energy lost.
The factor of $a/L$ appears because in this limit,
the particles remain tightly packed, and
rarely change places.  The ratio between $D_{pp}$ and $D_{pw}$ will be equal
to the ratio between the number of particle-particle contacts and the number
of particle-wall contacts.  Geometrical considerations show that these ratios
are of order $a/L$.

In the limit of $\upsilon\gg1$, Eq.~(\ref{eq:Dpw}) becomes [after using
Eq.(\ref{eq:KumaranDpp}) for $D_{pp}$] 
$D_{pw} \sim (1-r_p^2) N \bar E_x^{3/2} / (m^{1/2}L)$.  Then, realizing
that the average energy lost in particle wall collisions is 
$\langle E_w \rangle \sim (1-r_w^2) \bar E_x$ (note that $r_w=r_p$), 
and that the typical $x$ velocity will be $U_x=(\bar E_x/m)^{1/2}$, 
Eq.~(\ref{eq:Dpw}) becomes 
\begin{equation}
D_{pw} \sim N \langle E_w \rangle (U_x/L).
\label{eq:Dpwscaling}
\end{equation}
The similarity between this equation and Eq.~(\ref{eq:Dppscaling}) 
for $D_{pp}$ permits the following interpretation: in the large $h$ 
limit, the particles are bouncing back and forth between the two side
walls.  The factor $U_x/L$  gives the frequency of collisions with
the wall, so that $\langle E_w \rangle (U_x/L)$ gives the rate of
energy loss per particle.  There are $N$ particles, so a factor of
$N$ appears in Eq.~(\ref{eq:Dpwscaling}).

\begin{figure}[ht]
 \epsfig{file=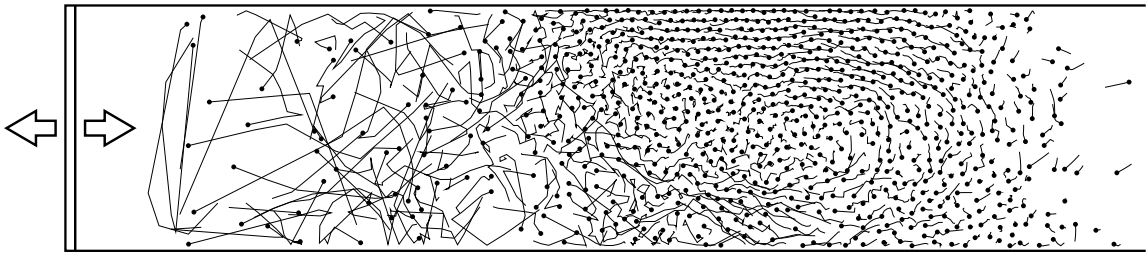,height=15.5cm,clip=,angle=270}
\caption{Streaklines showing the motion of each particle during one
cycle of the wall vibration.  The solid dot shows the position of
the particle at the beginning of the wall vibration, and the line shows
its motion during the cycle.  The simulation has $N=800$, $\rp=\rw=0.95$,
$\Bp=\Bw=-1$, $L=50$, $g=1$, $\tau=1$, and $V=15.5$ with wave form $B$.
The figure is horizontal, the vibrating bottom is on the left.}
\label{fig:convect}
\end{figure}

The deviation of the large $H$ simulations from Eq.~(\ref{eq:Dpw})
is probably due to their different mass distribution.  These simulations
differ from the others because they form a dense plug of particles
suspended by a dilute ``hot'' region.  In contrast, 
the others have
a density maximum near the bottom, similar to a normal atmosphere.
An additional curiousity is convection which appears in the large $N$
simulations (solid circles in Fig.~\ref{fig:Dpw}).
Fig.~\ref{fig:convect} shows the motion of all the particles
during one cycle for a $N=800$ simulation, revealing a
circulation of particles within the plug.  When $V$ is decreased
below a critical value, the circulation ceases, and this critical
value of $V$ corresponds to the break in the $N=800$ curve (solid circles)
in Fig.~\ref{fig:Dpw}.  We have not studied this convection in
detail.

A further complication arises
when the walls are rough ($\Bw\neq-1$).  One must decide whether the walls
move with the vibrating bottom or not; i.e.~whether the bottom is a piston
moving vertically between the stationary walls, or whether the
particles are contained in a box which is shaken.  If the side walls
also move, then energy can be input at the side walls as well as at 
the bottom (i.e.~$D_{pw}$ can be negative) \cite{luding95b}.  In this
paper, we always consider the side walls to be stationary.

\subsection{Gravitational Potential energy}

\begin{center}
\begin{figure}[ht]
 \epsfig{file=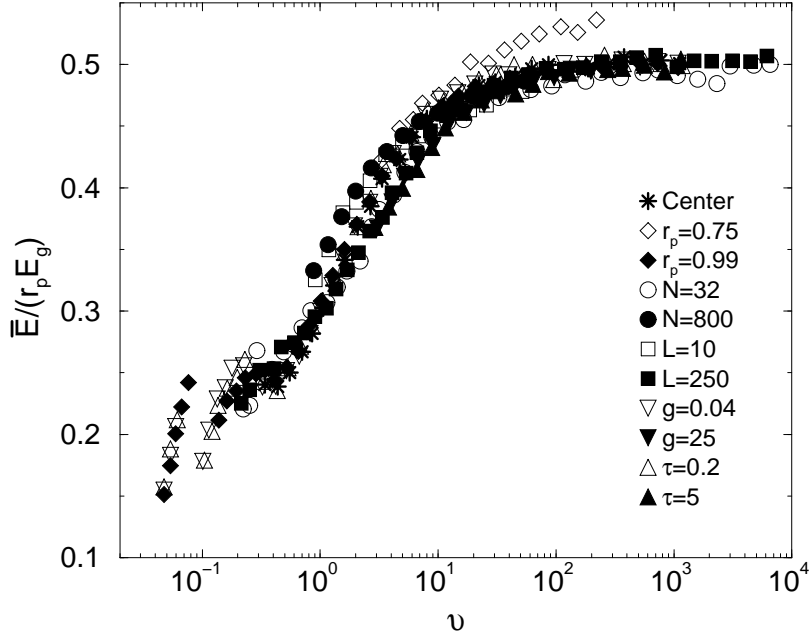,width=12cm,clip=,angle=0}
\caption{The ratio of translational kinetic energy $\Et$ to gravitational
kinetic energy $\Eg$ scaled by $1/\rp$ and
plotted against $\upsilon\equiv(h-h_0)/(aH) = L(h-h_0)/(2Na^2)$,
for the simulations in Fig.~\ref{fig:Dpp}.}
\label{fig:PEKE}
\end{figure}
\end{center}

As shown in Sec.~\ref{sec:Dpp}, $D_{pp}$ is a known function of the
system parameters, the energy and $h-h_0$.  We would like to eliminate
this dependence on $h-h_0$ before combining our results for $D_{pp}$,
$P_b$, and $D_{pw}$.  In order to do this, we plot in
Fig.~\ref{fig:PEKE} the ratio $\Et/(\rp\Eg)$ [where $\Eg \equiv
mg(h-h_0)$ is the gravitational potential energy per particle].
This scaling successfully collapses the data onto a single curve
reminiscent of Fig.~\ref{fig:Dpp}.  We note that $h$ must be
carefully calculated; if $h$ is taken to be the height above the lowest
position of the floor instead of the height above the average position of
the floor, the simulations with $g\tau^2=25$ fall off the curve.

Gravitational energy and kinetic energy are often assumed to be
equivalent.  Fig.~\ref{fig:PEKE} should be a cautionary note.
The dependence of $\Et/\Eg$ on the system parameters is quite
complicated, and has not been theoretically investigated.

\section{Summary and Test}
\label{sec:grail}

To summarize and test the formulas for $D_{pp}$, $P_b$, and $D_{pw}$
presented in the previous sections,
we calculate the predicted value of $\Et$ for a set of simulations
from a previous paper \cite{luding95b}.  
Using Eqs.~(\ref{eq:Dpprotate}), (\ref{eq:Pwall})
and (\ref{eq:Dpw}) to rewrite the energy
balance, Eq.~(\ref{eq:EnergyBalance}), as an equation in terms of
the system parameters and $U=(\Et/m)^{1/2}$, gives
\begin{equation}
gV \exp (-\alpha U/V) = 
	 C_{pp}{1-{\rp^*}^2\over1+\rp}gHU
	+ C_{pp}{1-{r_w^*}^2\over1+r_w}gHU C_{pw}(\upsilon)
	  \left({a \over L}\right)\left({\bar E_x\over\bar E}\right)^{3/2},
\label{eq:grail}
\end{equation}
where $C_{pp} = \min(0.30 \log[h-h_0]+1.35, \sqrt{2\pi})$, and
$C_{pw}=1.45+0.905\upsilon$.  In writing down Eq.~(\ref{eq:grail}), we
have assumed that Eq.~(\ref{eq:reff}) holds for the particle-wall restitution
coefficients.  To close Eq.~(\ref{eq:grail}), expressions for $(h-h_0)/a$
and $\bar E_x$ need to be supplied.  We assume $(h-h_0)/a=2\bar E/(mga)$
and $\bar E_x=\bar E$, which are true in a gas of elastic 
hard spheres under gravity ($r_p=1$, $\beta_p=\pm1$, $V=0$).
These assumptions hold approximately for granular media, except in some
extreme cases. 

\begin{center}
\begin{figure}[ht]
\epsfig{file=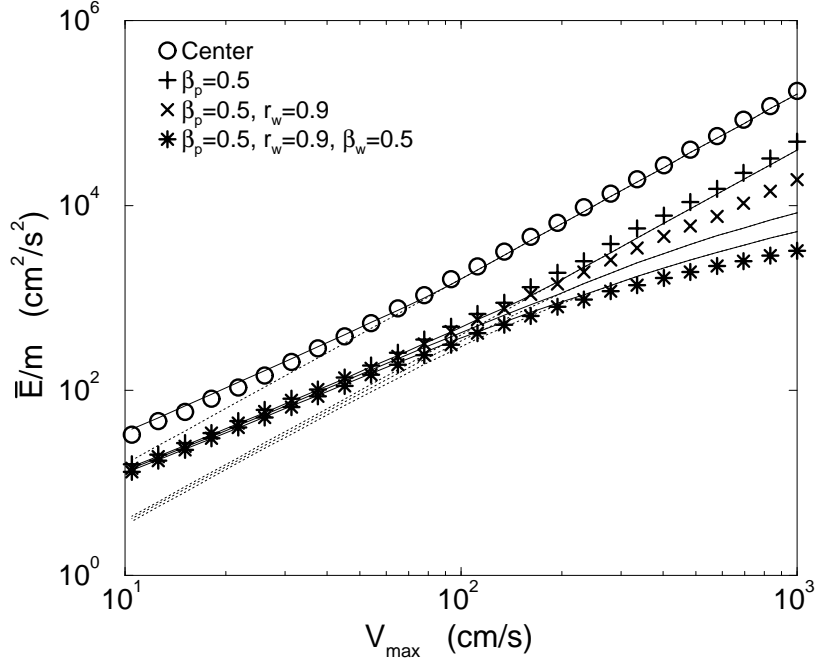,width=12cm,clip=,angle=0}
\caption{A test of the theoretical results from Eq.~(\ref{eq:grail}). 
The central simulations
have $N=50$, $a=0.05\ \text{cm}$,
$L=20a$, $\rp=0.9$, $g=981\ \text{cm/s}^2$, $\tau=0.01\ \text{s}$, $\rw=1.0$,
and $\Bp=\Bw=-1$, with wave form $C$.
The other simulations are the same, except for the parameters given
in the figure legend.  The solid lines give the results of the theory
presented in Eq.~(\ref{eq:grail}).
The dotted lines give the theory without taking into account the
logarithmic dependence of $C_{pp}$ on $h-h_0$, 
i.e., with $C_{pp}  = \protect\sqrt{2\pi}$. }
\label{fig:grail}
\end{figure}
\end{center}

We test this equation against simulations in Fig.~\ref{fig:grail}.  
The parameters of the central simulations are $N=50$, $a=0.05\ \text{cm}$,
$L=20a$, $\rp=0.9$, $g=981\ \text{cm/s}^2$, $\tau=0.01\ \text{s}$, $\rw=1.0$,
and $\Bp=\Bw=-1$, with wave form $C$.  
These parameters were chosen to mimic the
simulations in Fig.~11 of Ref.~\cite{luding95b}.
There remain three
differences between these simulations and those of Ref.~\cite{luding95b}.
First, Ref.~\cite{luding95b} uses
a more complicated collision rule,
where the tangential restitution coefficient varies between $-1$ and
an upper limit $\beta_{0}$, depending on the impact parameter.  
In our simulations, the tangential restitution coefficient is
the same for all collisions.
Secondly, Ref.~\cite{luding95b} used slightly polydisperse
spheres, whereas we use monodisperse spheres.  Finally, we replace
the sinusoidal wave form of Ref.~\cite{luding95b} by wave form $C$.
In spite of these differences, the simulations presented here show the
same behavior as those of Ref.~\cite{luding95b}.

We show two versions of the theory, one which takes into account the 
logarithmic dependence of $C_{pp}$ (the solid lines) and the other
which does not (the dotted lines).   The solid lines reproduce the
observed dependence $\bar E \sim V^{3/2}$ for 
$V_{\text{max}}<100\ \text{cm/s}$, 
whereas the dotted lines show $\bar E \sim V^2$.  Thus, the puzzling
scaling observed in previous work \cite{luding95b,warr95} is due
to the logarithmic dependence of $C_{pp}$ on $h-h_0$.  This dependence
does not yet have a theoretical explanation.

For $V_{\text{max}}>100\ \text{cm/s}$, there is a significant
disagreement for the simulations
with dissipative walls ($\rw\neq1$), which can be traced back to the
failure of the assumption $\bar E_x= \bar E$.

\section{Conclusions}

This paper has studied the two dimensional vibrated granular media,
using the energy balance Eq.~(\ref{eq:EnergyBalance}) to organize
our investigation.  Reviewing existing theories, we find the
particle-particle dissipation is well understood in the dilute limit.
We were able to show that Eq.~(\ref{eq:KumaranDpp}) is very accurate
in the absence of side walls.  This result was extended
to deal with rough rotating particles with a constant tangential
restitution coefficient.  More work is needed to account for
the more realistic situation of variable tangential restitution.
In the dilute limit, particle-particle dissipation is well
accounted for by existing theories, but in
the dense limit, it shows an
unexplained dependence on the height of the center of mass above
the vibrating bottom, $h-h_0$, well approximated by a logarithm.
It is this departure from the dilute theory which accounts for the
$E \sim V^{3/2}$ scaling observed in previous work.  The cause of this
departure is not yet known: it is not due to a change in density,
since it depends only on the height of the center of mass.  Nor can
it be explained by waves propagating upwards from the bottom.
Furthermore, the ratio between potential and kinetic energy, $\Et/\Eg$,
 can vary by almost a factor of $3$.  This variation is not
understood, and hampers the ability of our theory to predict
$h-h_0$.

The energy input by the vibrating plate is also well known for
the special driving wave form $B$ in Fig.~\ref{fig:system}(b).
For more conventional wave forms, the energy input is well approximated
by the exponentially decaying function of $U/V$ shown in Eq.~(\ref{eq:Pwall}).

To our knowledge, this is the first paper to treat particle-wall dissipation
in detail.  Our theory accurately predicts energy losses at the wall, but
requires knowing the difference between the horizontal and vertical
kinetic energies.  A more complete theory would have to estimate this
difference from the system parameters.

We acknowledge the generous support of the Alexander von Humboldt-Stiftung
and the DFG, SFB 382 (A6).

\appendix
\section{The Collision rule}
\label{sec:Collrule}

Consider a collision between two particles of radius $a$.  Let $\bv_1$ be
the translational velocity of the first particle, $\bbox{\omega}_1$
its angular velocity and $\bbox{r}_1$ its position at the time
of contact.  The quantities $\bv_2$, $\bbox{\omega}_2$, and
$\bbox{r}_2$ are the analogous quantities for the second particle.
Then the relative velocity at the point of contact is
\begin{equation}
\bv_c = \bv_1 - \bv_2 - a(\bbox{\omega}_1 + \bbox{\omega}_2) \times \n.
\end{equation}
Here, the unit vector
$\n \equiv {\left ( \bbox{r}_1 - \bbox{r}_2 \right ) /
                \left| \bbox{r}_1 - \bbox{r}_2 \right|}$
points along the line connecting the centers of the
colliding particles, from particle 2 towards particle 1.
The change in the normal component of $\bv_c$
is parameterized by the ``coefficient of normal restitution'' $\rp$, so that
$\bv_c' \cdot\bbox{\hat n} = - \rp (\bv_c \cdot\bbox{\hat n})$,
where the prime denotes the velocity after the collision.
When $\rp = 1$, energy is conserved, and energy dissipation requires
$0 \le \rp < 1$.
The coefficient of tangential restitution $\Bp$ is defined analogously,
i.e. $\bv_c' \times\bbox{\hat n} = - \Bp (\bv_c \times\bbox{\hat n})$.
Energy is conserved for $\Bp=-1$ (perfectly smooth surfaces) and for
$\Bp=1$ (perfectly rough surfaces).
In the first case the $\bbox{\omega}_i$ have no effect on outcome of
the collision, and do not change during the collision.
Energy is dissipated when $\Bp$ lies between these two extremes.

From the definitions of $\rp$ and $\Bp$ and the assumption that
the interaction takes place only at the point of contact, 
it is possible to derive the collision rules 
\begin{eqnarray}
\bv_{1,2}' &=& \bv_{1,2} \mp {1+\rp\over 2} \bv_n \mp
    {q(1+\Bp) \over 2q+2} (\bv_t+\bv_r) , {\text{~and}}\cr
a\bbox{\omega}_{1,2}' &=& a\bbox{\omega}_{1,2} + {1+\Bp \over 2q+2} 
                \left[\bbox{\hat n}\times(\bv_t + \bv_r)\right],
\label{eq:collrule}
\end{eqnarray}
where the equation for particle 1 (2) takes the $-$ ($+$) sign in the
top line above.  To derive Eq.~(\ref{eq:collrule}) we used momentum
conservation and the definitions
\begin{eqnarray}
\bv_n &\equiv& \left [ (\bv_1 - \bv_2) 
    \cdot \bbox{\hat n} \right ] \cdot \bbox{\hat n} , \cr
\bv_t &\equiv& \bv_1 - \bv_2 - \bv_n, \quad \text{and} \cr
\bv_r &\equiv& - a(\bbox{\omega}_1 + \bbox{\omega}_2) \times \bbox{\hat n}.
\end{eqnarray}
Here, $\bv_n$ is the normal component of $\bv_c$, $\bv_t$
is the tangential component due to  translation, and $\bv_r$ is
the tangential component due to rotation.  Note that $\bv_c =
\bv_n + \bv_t + \bv_r$.

The change in translational energy is
\begin{eqnarray}
\Delta \bar E = - Q v_n^2 - S
  \left[C_{t1} v_t^2 + C_{t2} (\bv_t\cdot\bv_r) - C_{t3} v_r^2\right],
\label{eq:deltaEt}
\end{eqnarray}
with the positive prefactors $Q \equiv m(1-r^2)/4$, 
$S \equiv mq(1+\Bp)/[4(1+q)^2]$, 
and the constants $C_{t1} \equiv 2+q(1-\Bp)$,
$C_{t2} \equiv 2-2q\Bp$ and $C_{t3} \equiv q(1+\Bp)$.
Likewise, the change in rotational energy is
\begin{eqnarray}
\Delta\overcirc E = - S
  \left[- C_{r1} v_t^2 + C_{r2} (\bv_t\cdot\bv_r) + C_{r3} v_r^2\right],
\label{eq:deltaEr}
\end{eqnarray}
where the constants are $C_{r1} \equiv (1+\Bp)$,
$C_{r2} \equiv 2(q-\Bp)$, and $C_{r3} \equiv 2q+1-\Bp$.  
Note that the $C$ are positive (only $C_{r2}$ can also
be negative) so that the signs in Eqs.\ (\ref{eq:deltaEt}) 
and (\ref{eq:deltaEr}) indicate the direction
of energy transfer between the degrees of freedom.

Eqs.~(\ref{eq:deltaEt}) and~(\ref{eq:deltaEr}) can be added
together to give 
\begin{equation}
\Delta E = - Q v_n^2 - S(1+q)(1-\Bp)(\bv_t+\bv_r)^2.
\end{equation}
In this paper, we need to know the average energy lost per collision.
Using angle brackets to denote averages over collisions, we have
\begin{equation}
\langle\Delta E\rangle = - Q \langle v_n^2\rangle
		- S(1+q)(1-\Bp)\langle(\bv_t+\bv_r)^2\rangle.
\end{equation}
Assuming that the particles' velocities are distributed according to a
Maxwellian velocity distribution, and that their postions and velocities
are uncorrelated gives
\begin{equation}
\langle v_n^2\rangle = 8\Et/m, \quad
\langle(\bv_t+\bv_r)^2\rangle = 4(\Et+\Er)/m,
\end{equation}
in two dimensions. \cite{mcnamara97b}
Thus the total energy lost during one collision is
\begin{equation}
\langle\Delta E\rangle = - 2 (1-\rp^2) \Et
		- {q(1-\Bp^2)\over 1+q}\left(\Et+\Er/q\right).
\label{eq:DeltaEappendix}
\end{equation}

\bibliographystyle{prsty}

 
\end{document}